# Modulating dislocation reactions through preferential hydrogen segregation in bcc metals

Jie Hou[1], Ducheng Peng[1], Xiang-Shan Kong[2], Huiqiu Deng[3], Wangyu Hu[1], Cheng Chen[4*], Jun Song[5*]

1. *College of Materials Science and Engineering, Hunan University, Changsha, 410082, China*
2. *State key Laboratory of Advanced Equipment and Technology for Metal Forming, Shandong University, Jinan, Shandong 250061, China*
3. *School of Physics and Electronics, Hunan University, Changsha 410082, China*
4. *School of Aeronautics, Northwestern Polytechnical University, Xi'an, Shaanxi 710072, China*
5. *Department of Mining and Materials Engineering, McGill University, Montreal, Quebec, H3A 0C5, Canada*

**Abstract**

The interaction between dislocations is fundamental to plastic deformation, work hardening, and defect accumulation. While extensive research has focused on the impact of solutes on individual dislocations, how solutes affect dislocation-dislocation reactions remains largely unexplored. Here, using atomistic simulations of iron as a model bcc system, we demonstrate that hydrogen solutes enable two <111>/2 screw dislocations to react and form a <001> edge dislocation junction, a process that is otherwise unfavorable in hydrogen-free environments. This phenomenon arises from the preferential segregation of hydrogen around the <001> dislocation, which reduces the energy of the reaction product. The resulting <001> dislocation demonstrates remarkable stability and transforms into a <001> vacancy-type dislocation loop under strain. These vacancy-type dislocation loops can accumulate during continuous deformation and dislocation reactions, serving as precursors for the initiation of structural damage, such as cracking and blistering. Our findings highlight the pivotal role of hydrogen in dislocation reactions, uncover a novel defect accumulation mechanism crucial for interpreting recent experimental observations, and represent a significant advance in understanding hydrogen-induced damage in bcc metals.

**Keywords:** Dislocation reaction; hydrogen induced damage; atomistic simulation; bcc metals;

* Correspondence should be addressed to cheng.chen@nwpu.edu.cn or jun.song2@mcgill.ca

**Graphical abstract**

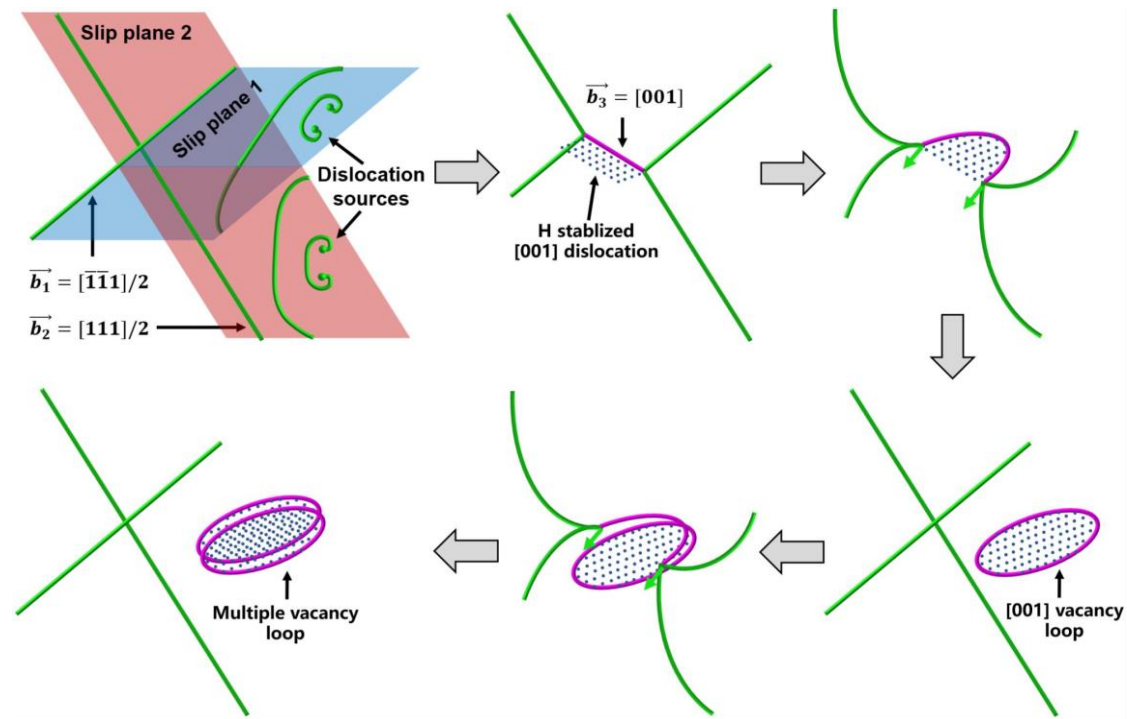

**Introduction**

Dislocations are linear crystallographic defects present in nearly all metallic materials and serve as the primary carriers of plastic deformation [1]. It is well established that solute atoms can interact with dislocations, thereby influencing the mechanical performance of metals. Such interactions can be beneficial, as solutes may increase the critical stress required for dislocation slip, a phenomenon known as solute strengthening [2]. Conversely, these interactions can also have detrimental effects, potentially compromising plasticity or facilitating crack and blister formation [3-5]. Among different solute-dislocation interactions in metals, the interplay between hydrogen (H) and dislocation has received significant attention, owing to hydrogen's ubiquitous presence in service environments and its potential to jeopardize the toughness (thus the safety) of metal components.

In the pursuit of understanding H interactions with dislocations, extensive research has led to the development of numerous theories and models. One of the earliest and most influential is the classic Cottrell atmosphere model [6], which predicts the segregation and dragging of interstitial solutes around dislocations. This model was later refined to incorporate more realistic representations of dislocation core structures [7]. More recently, atomistic simulations have been performed to study H-dislocation interaction, demonstrated H suppressed dislocation motion that consistent with solute dragging theory [8], and revealed H suppressed dislocation emission from crack tip which promotes cleavage fracture [3,9,10]. In addition to these H suppression effects on dislocation activity, there are also mechanisms proposed to argue that H may facilitate dislocation nucleation and slipping, such as H enhanced local plasticity [11], defactant theory [12-15], H modified kink-pair formation and migration [16-18], to name a few. The above are only a small set of the long list of great research studies that together build the valuable knowledge base for the complex puzzle of H effect on dislocation and plasticity.

Yet, despite those numerous efforts and scientific advances made, critical gaps remain in our understanding of H-dislocation interactions. Particularly worth noting is that most previous theoretical and modeling studies were focused on individual or few non-intersecting dislocations [8,16,19-29], which is an oversimplification for the actual scenario in materials containing complex dislocation networks populated with dislocation junctions that form from dislocation-dislocation reactions [30-34]. As a result, the role of H in dislocation junction formation remains absent. With dislocation junctions being major pinning sites for dislocation motion and significantly contribute to strain hardening, this necessarily renders a crucial knowledge deficit toward understanding how H influences plastic deformation in realistic materials.

In recent years, the need to address such deficit has become increasingly pressing, particularly in light of several experimental observations [4,5,35-38]. For instance, Guo et al. [5] observed the nucleation of H blisters at dislocation tangles in W, along with the presence of <001> dislocation junctions that typically formed from the reaction of two <111>/2 dislocations. Similarly, Chen et al. [4], found a connection between H blister and <001> dislocation, explicitly showing that crack-shaped intra-granular H blisters initiate along {001} planes. Cho et al. [35] also showed substantial H induced {001} cleavage in Fe, with evident dislocation slipping bands intersecting near the cleavage plane. These findings clearly highlight the key role played by dislocation-dislocation reaction and <001> dislocation junctions in H induced damages. In addressing these experimental observations, theoretical efforts were also made recently to evaluate the effect of H on <001> dislocation junctions [22,39]. Yet these efforts were either still limited to individual dislocation lines without considering dislocation reactions [22], or based on linear elastic approximation with H-H interaction ignored [39], which is known to greatly underestimate H-dislocation interactions (especially the <001> ones [22]). Overall, we still know very little about how H affects dislocation reactions and network evolution in metals.

Our present work aims to directly address these issues. Using bcc Fe as a model system, with an interatomic potential that modified to accurately describes H-H interactions and clustering around dislocations [3,22,40], we carried out systematic hybrid molecular dynamics (MD) and grand canonical Monte Carlo (GCMC) simulations. Our results demonstrate that hydrogen preferentially segregates around <001> edge dislocations, significantly reducing their free energy. This segregation facilitates the formation of <001> dislocation junctions from the reaction between two <111>/2 screw dislocations, a process that is typically unfavorable in hydrogen-free environments. The <001> junctions formed can evolve into <001> vacancy-type dislocation loops under shear loading, and with continuous dislocation reactions, these loops can accumulate to form multilayer vacancy loops, which serve as precursors for blistering or cracking. Our discoveries reveal the important role played by H in dislocation reaction and the formation of <001> edge junction and vacancy loops. These insights provide atomistic details critical for interpretating recent experimental observations, offering a new perspective on dislocation reaction and network evolution in bcc metals under H rich environments.

**Results**

**Dislocation reaction in pristine Fe**. In this work, we focus on the reaction between two <111>/2 screw dislocations, which are typically the dominant dislocation type in bcc metals [34]. As illustrated in Fig. 1a-1b, when two of these dislocation intersect, they can form a <001> type dislocation junction following the Burgers vector summation $[111]/2 + [\bar{1}\bar{1}1]/2 = [001]$ [30]. According to the line tension model [41], such dislocation reaction is feasible when it leads to a reduction in the total line energy of the system, namely:

$$\Delta E = E_3 - (E_1 \cos\theta_1 + E_2 \cos\theta_2) < 0, \qquad (1)$$

where $E_1 = E_2$ represents the line energy for the <111>/2 screw dislocation, $\theta_1 = \theta_2 = 35.3°$, and $E_3$ is the line energy for the <001> edge dislocation junction.

To evaluate the feasibility of the above junction formation process, we first calculated the dislocation line energies in bcc Fe using MD simulations. The calculated line energies of various dislocations as functions of the cutoff radius are shown in Fig. 1c. A clear linear relationship between the line energy and the logarithm of the cutoff radius was observed, consistent with the predictions from linear elastic theory [1]. Upon closer inspection of the energy values, we find that the line energy of a single <001> edge dislocation is generally higher than that of two <111>/2 screw dislocations. Using these line energy data, we further calculated $\Delta E$ based on Eq. (1). As shown in Fig. 1c, $\Delta E$ is predominantly positive, except for a few points near the dislocation core, indicating such junction formation process is energetically unfavorable. This behavior is also observed in other bcc metals, including V, Nb, Ta, Cr, Mo, and W (see supplementary information for details), suggesting that the formation of <001> edge dislocation junctions from two <111>/2 screw dislocations is generally unfavorable in pristine bcc metals.

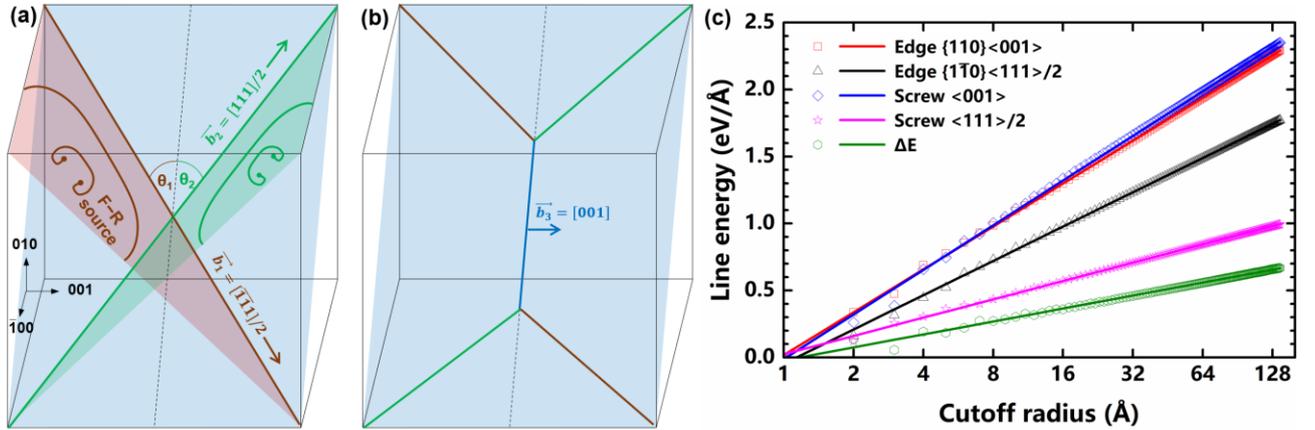

**Figure 1. Reaction between two <111>/2 screw dislocations.** (a-b) Schematic representation of the intersection between a $[111]/2$ screw dislocation and a $[\bar{1}\bar{1}1]/2$ screw dislocations, leading to the formation of a $[001]$ edge dislocation. (c) Line energy of various dislocations as a function of cutoff radius. The change in energy ($\Delta E$) for junction formation, calculated using Eq. (1), is also shown. Symbols represent results from MD simulations, and lines indicate the corresponding logarithmic fittings.

**Strong binding between H and <001> edge dislocation junction.** Note the above discussion pertains to the behavior of pure Fe without considering the effects of H. While in our previous work [22], we demonstrated that H strongly favors tension along <001> directions and tends to cluster around <001> edge dislocations. Given that hydrogen can significantly reduce dislocation line energy [21], this preferential clustering may greatly stabilize <001> edge dislocations, facilitating the formation of otherwise unfavorable <001> junctions from two <111>/2 screw dislocations. This, in turn, could drastically alter the mechanical behavior of Fe in H-rich environments.

To verify this hypothesis, additional hybrid MD+GCMC simulations were conducted. These MD+GCMC simulations enable the modeling of systems with arbitrary far-field H concentrations (or corresponding H chemical potentials) at thermodynamic equilibrium, allowing for the study of H effects on dislocations. Fig. 2a presents the distribution of H atoms around different dislocations at varying H concentrations. It is evident that the <001> edge dislocation attracts a significant amount of H on its tensile side, forming a pronounced H cluster that extends nanometers beneath the slip plane at high H concentrations. In contrast, the clustering of H around other dislocations is less pronounced, with H atoms confined within a few atomic layers beneath the slip plane of <111>/2 edge dislocations, and H localized only at the core regions of two screw dislocations. This preferential H clustering stems directly from the tensile stress field around the <001> edge dislocation, which has been shown to attract H more effectively than tension along other directions, such as <111> (as detailed in Ref. [22]).

This preferential H clustering was further quantified by examining the H excess for different dislocations, defined as the difference in the number of H atoms between a defected system and a defect-free system of the same size. The calculated H excess is shown in Fig. 2b. In addition to the pronounced preferential H clustering at the <001> edge dislocation (1–3 orders of magnitude higher than others), several noteworthy observations emerge. First, the <111>/2 screw dislocation consistently exhibits the lowest H excess across all H concentrations, likely due to its lowest H binding energy of 0.261 eV, compared to 0.415 eV for the <111>/2 edge dislocation, 0.425 eV for the <001> screw dislocation, and 0.588 eV for the <001> edge dislocation. Second, the slope of the H excess curve for the <001> screw dislocation is significantly smaller than for the other dislocations, likely due to the absence of long-range elastic interactions. This results from the fact that shear stress along <001> directions does not couple with the tetragonal misfit volume of H [39]. In contrast, for the <111>/2 screw dislocation, shear stress along <111> directions can produce non-zero elastic interactions with H. Therefore, despite both being screw dislocations, the <111>/2 screw dislocation is more sensitive to H concentration than the <001> screw dislocation.

Based on the observed preferential H clustering behavior, we anticipate distinct impacts of H on the line energy of different dislocations. This can be quantified by calculating the binding free energy of H to various dislocations, which reflects the change in free energy due to H binding. As shown in Fig. 2c, H exhibits a strong binding affinity for the <001> edge dislocation, with the binding free energy increasing significantly with H concentration, and markedly higher than that for the other dislocations. Notably, at high H concentrations, the binding free energy for the <001> edge dislocation can approach or even exceed the line energy shown in Fig. 1b. This suggests a substantial

H stabilization effect on the <001> edge dislocation, highlighting the potential for H to enable the formation of otherwise unfavorable <001> edge dislocation from two <111>/2 screw dislocations. Although our simulations considered only {110} slip planes for edge dislocations, we expect the influence of slip planes on these results to be minimal, as the stress field of a dislocation plays a more crucial role in attracting H than its core structure.

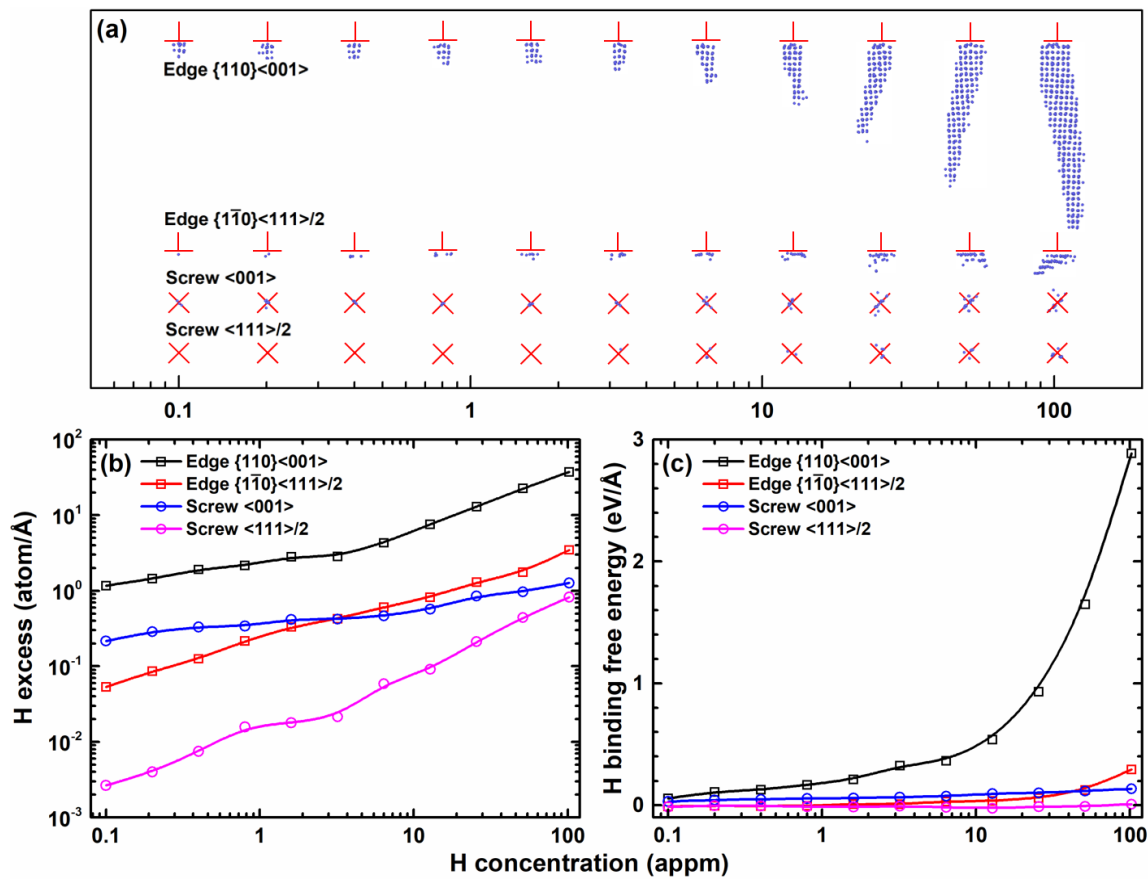

**Figure 2. H segregation around dislocation lines in Fe at varying far-field H concentrations.** (a) Representative configuration of H atoms around different dislocations; Fe atoms are not shown for clarity. (b) Number of excess H atoms around different dislocations. (c) H binding free energy to different dislocations.

**H enabled <001> edge dislocation junction formation.** In light of the above findings, we next directly investigated the formation of H-enabled <001> edge dislocation junction. Initially, two <111>/2 screw dislocations were introduced in neighboring ($1\bar{1}0$) planes, and a small shear strain was applied to induce a metastable crossed state configuration [42]. Following this, we conducted a two-step GCMC H charging process: first with the shear strain applied, and subsequently with the shear strain removed (additional details can be found in supplementary information). After reaching equilibrium, the resultant dislocation structures under various far-field H concentrations are illustrated in Fig. 3.

For a low H concentration of 1.6 appm, we observed that the two <111>/2 screw dislocations bow out and move away from each other, indicating a repulsive interaction. This behavior suggests that despite the dislocations being initially forced together by shear loads, they revert to their original positions once the shear is removed. This observation aligns with the behavior in H-free conditions, implying that 1.6 appm of H is insufficient to influence the dislocation interaction. In contrast, at a higher H concentration of 6.4 appm, the two <111>/2 screw dislocations do not spring back but instead maintain a relatively stable crossed state even after the shear is unloaded. This indicates a stronger attraction between the dislocations due to the presence of H. With further increases in H concentration to 25.6 and 102.4 appm, a distinct {110}⟨001⟩ edge-type dislocation junction becomes apparent by the end of the simulations, with a significant accumulation of H atoms at the tension side of the junction. It is noted that these H clusters are smaller compared to those shown in Fig. 2a. This difference is expected, as the tensile field of a finite junction segment is weaker than that of an infinitely long dislocation line (see supplementary information).

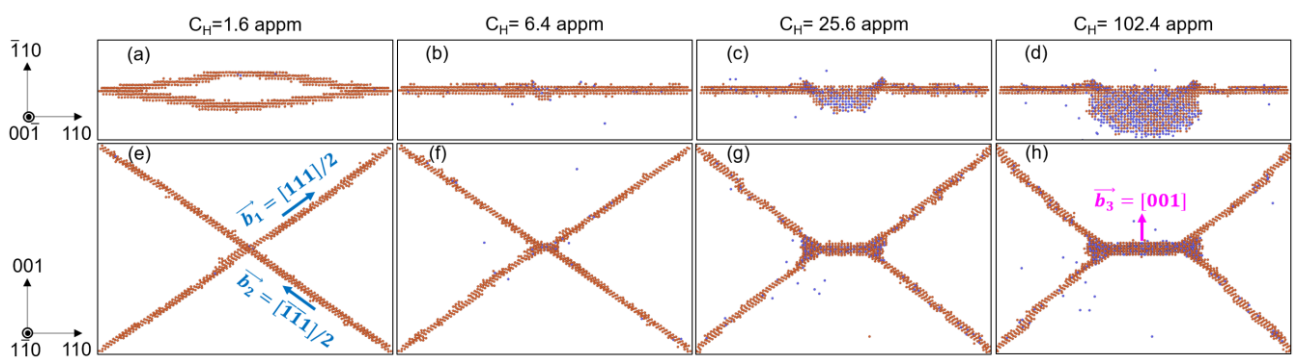

**Figure 3. H-enabled <001> edge junction formation from two <111>/2 screw dislocations.** Orange atoms represent Fe (only non-bcc Fe atoms are visualized), and blue atoms are H. (a-d) Top views from the $[00\bar{1}]$ direction under different far-field H concentrations. (e-f) Corresponding front views from the $[1\bar{1}0]$ direction.

**H induced <001> vacancy loop formation and accumulation.** The above findings support our hypothesis that H can enable the formation of <001> edge junctions from two <111>/2 screw dislocations. However, these observations were mostly static. To understand how the <001> edge junction affects dislocation dynamics during plastic deformation, we conducted MD+GCMC simulations with shear loads to drive continuous dislocation movement. We focused on a H concentration of 102.4 appm, starting with the structure shown in Fig. 3d and 3h. A shear load was applied to exert the same resolved shear stress on the two <111>/2 screw dislocations, driving them to move in the $[1\bar{1}0]$ direction (see supplementary information). The simulation results are presented in Fig. 4a-e. Contrary to the junction unzip mechanism often observed in previous studies [30,39,43-45], where <111>/2 dislocation arms bow out to reduce junction length and eventually annihilate it, we observed that the movement of <111>/2 screw dislocations did not lead to complete unzipping of the junction. Instead, the dislocations bypassed the junction sequentially, leaving behind a vacancy-type <001> dislocation loop (referred to as a <001> vacancy loop). This loop, approximately rectangular with a side length of ~40 Å, contains a significant amount of clustered H atoms within the loop region.

Considering that <001> dislocations are generally sessile and vacancy loops in bcc metals usually do not unfault as they do in fcc metals [46], we expect the <001> vacancy loop to remain relatively stable during plastic deformation. This stability suggests that the loop could act as a persistent obstacle, potentially blocking and interacting with other dislocations during subsequent deformation. To investigate this, we set up a model with a pre-existing rectangular <001> vacancy loop and inserted two <111>/2 screw dislocations at two neighboring atomic layers adjacent to the loop. We then performed GCMC H charging to establish equilibrium, followed by applying shear loading and GCMC H charging in the same manner as described previously.

The simulation results with a pre-existing <001> vacancy loop, shown in Fig. 4f-4j, reveal a similar dislocation evolution to that observed in Fig. 4a-4e, although some vacancy trails were noted due to jog dragging. Specifically, two <111>/2 screw dislocations sequentially bypass the junction, and no annihilation of the <001> dislocations was observed. After shear loading, the <001> vacancy loop transformed into a <002> vacancy loop, now decorated with a thicker H cluster. To extend this investigation, we further performed simulations with a pre-existing <002> vacancy loop. As illustrated in shown in Fig. 4k-4o, <002> vacancy loop evolved into a <003> vacancy loop following further interaction with two <111>/2 dislocations. These results suggest that the initial <001> vacancy loop can persist by continuously blocking and interacting with subsequent <111>/2 screw dislocations, leading to the formation of H-decorated multilayer <001> vacancy loops. Preliminary simulations suggest that vacancy loops could promote the opening of bulk lattice structures along {001} planes, potentially acting as precursors to lattice damage phenomena such as cracking or blistering. (see supplementary for details).

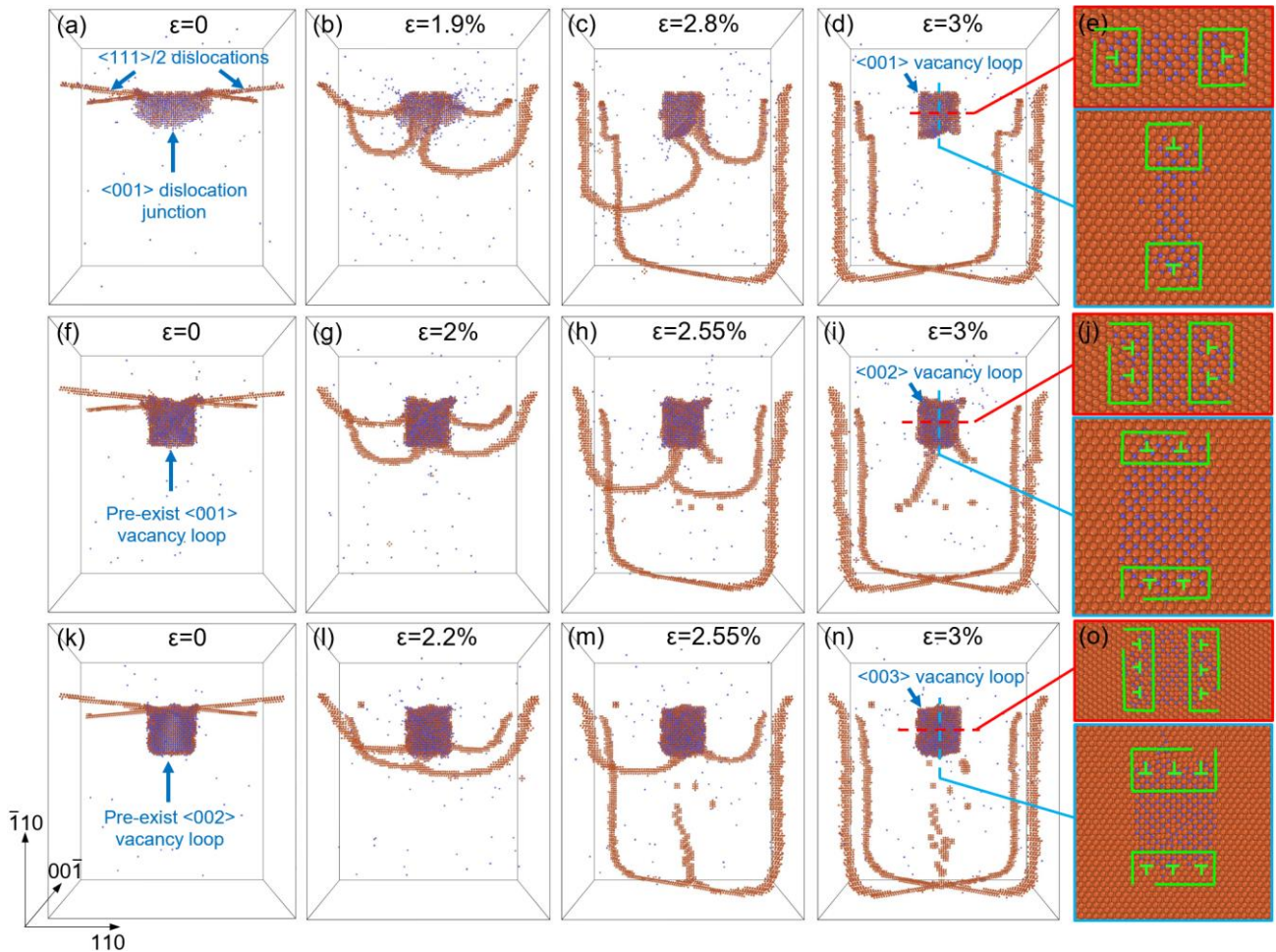
8

**Figure 4. Creation of vacancy dislocation loops from two <111>/2 screw dislocations**. Orange atoms are Fe and blue ones are H. (a-d) Show the structural evolution (only non-bcc Fe atoms are visualized) starting from the [001] dislocation junction demonstrated in Figs. 4d and 4h. (e) Displays cross-sections and Burgers vector analysis of the <001> vacancy dislocation loop created. (f-j) and (k-o) Demonstrate results from the same procedure, but starting with pre-existing <001> and <002> vacancy dislocation loops, respectively.

**Discussion**

The present study uncovers a novel mechanism by which H modifies the interaction between two <111>/2 screw dislocations, promoting the formation of <001> edge dislocation junctions and vacancy loops. While all the simulations presented above were conducted using Fe as a model bcc system, we posit that this mechanism is broadly applicable to other bcc metals. This is supported by our previous study, which demonstrated that preferential H clustering in the <001> tensile regions is consistent across several bcc metals, including Fe, W, Mo, and Cr [22]. In the current study, we also conducted benchmark simulations that directly show H can enable <001> edge dislocation junction formation in W (see supplementary information). Therefore, we expect the key findings of this work to be applicable to bcc transition metals in general.

Given that <111>/2 screw dislocations are the predominant dislocation type in bcc metals [34], the presence of H is expected to significantly influence microstructure evolution during plastic deformation. This effect could underpin several recent experimental findings, which have highlighted strong correlations between dislocation reactions and H-induced damage. For example, Guo et al. demonstrated that H blister nucleation in bcc W is directly linked to the formation of <001> edge dislocation junctions [5]. A similar connection was observed by Chen et al. [4], who explicitly showed that crack-shaped intragranular blisters nucleate on {001} planes. Cho et al. reported substantial H-induced {001} cleavage across martensitic laths, with dislocation slip bands intersecting near the cleavage plane [35]. With the main discovery of this work, i.e., H induced <001> junction and vacancy loop formation, these recent observations can now be elucidated at the atomic level. Furthermore, Deng et al. demonstrated the formation of <001> dislocations and loops following bending tests on H-charged samples [38], and they revealed H-restricted dislocation mobility near crack tips [37]. This behavior may result from H-enhanced dislocation pinning, where stable <001> edge junctions effectively pin two <111>/2 dislocations, thereby increasing flow stress (see Supplementary Information) and reducing dislocation mobility. Overall, these experimental results underscore the crucial role of dislocation reaction in H-charged bcc metals, providing robust support for our simulation results.

In summary, our atomistic simulations uncover a novel mechanism wherein H modulates the reaction between two <111>/2 screw dislocations in bcc Fe, facilitating the formation of otherwise unfavorable <001> edge dislocation junctions. This effect is attributed to the preferential aggregation of H around <001> edge dislocations, which reduces their line energy and promotes the junction formation. The <001> dislocation junction remain stable under shear loading, leading to the formation of <001> vacancy dislocation loops via the bypassing of two <111>/2 screw dislocations. These loops can accumulate locally with continued deformation and dislocation reactions, potentially serving as nucleation sites for H-induced damage such as blisters and cracks. The atomistic details revealed in this work are crucial for interpreting recent experimental observations and offer a new perspective on dislocation reactions and network evolution in bcc metals in H-rich environments.

**Methods**

**General setups in molecular dynamics simulations.** All simulations in this work were conducted using the Large-scale Atomic/Molecular Massively Parallel Simulator (LAMMPS) package [47], with visualization of results performed using the OVITO software [48]. We employed the embedded atom method (EAM) potential developed by Ramasubramaniam et al. [40] with modifications to the hydrogen-hydrogen attraction as described in Ref. [3]. This potential has been shown to accurately describe dislocation core structures (see Supplementary Information) and H clustering behaviors [22]. The canonical (NVT) ensemble was adopted in all simulation at a temperature of 300 K, regulated using a Nose-Hoover thermostat. The equations of motion were integrated with a timestep of 1 fs. A lattice constant of 2.86 Å, representing the average value for pressure-free Fe at 300 K, was used throughout the simulations. For static simulations without applied strain, the system was first relaxed for 1 ns to reach equilibrium, followed by an additional 1 ns of simulation to collect statistical data. In simulations involving shear loading, the fully relaxed system was subjected to a shear strain rate of 0.05/ns.

**Grand-canonical Monte Carlo simulations**. For simulations involving H, the MD methods were hybridized with GCMC simulations. This approach enables the establishment of thermodynamic equilibrium without requiring a full kinetic simulation of H redistribution, which is typically infeasible due to the large simulation box sizes and the long diffusion times required at low H concentrations.

In GCMC portion of the simulation, H atoms were randomly inserted or deleted according to the standard GCMC algorithm [49] to achieve a designated H chemical potential, $\mu$, within a constant volume (i.e., $\mu$VT). Once equilibrium is established, H atoms in different lattice environments (e.g., near the dislocation core or within bulk lattice) will correspond to the same chemical potential. A series of calculations using defect-free bcc Fe were first carried out with different $\mu$ values (details can be found our previous work in Ref. [22]), which gives us an Arrhenius type relation between $\mu$ and the far-field H concentration: $C_H = \exp(\frac{\mu + \mu_0}{k_B T})$, where $\mu_0 = 2.051$ eV, $k_B$ is the Boltzmann constant, and T = 300 K.

The MD and GCMC steps were alternated and iteratively executed in a specified proportion, consisting of 20 MD timesteps following $N \times 20$ GCMC attempts of H insertion or deletion, with N being the ratio between GCMC and MD steps. Here, $N$ represents the ratio between GCMC and MD steps, generally ranging from 0 to 10. The value of $N$ was chosen based on factors such as H concentration, system volume, and the difference between initial and equilibrium H distributions. It was determined on an ad hoc basis, with the guiding principle being that $N$ should be sufficiently large to ensure statistical convergence in H number and distribution. For H-free systems, $N = 0$ was used to perform pure MD simulations.

**Simulations of individual dislocations.** A cylindrical model with a radius of 150 Å and a thickness of approximately 20 Å was employed. Free surface boundary conditions were applied along the two radial directions, while the model remained periodic along its longitudinal axis. A dislocation line was constructed along the longitudinal axis using the Atomsk package [50], with atoms displaced according to anisotropic linear elastic theory [51]. To minimize boundary effects from the outer surface of the cylinder, atoms in the outer shell, with a radial thickness of 7 Å, were held fixed after displacement, while only atoms in the inner region were free to move. The line energy of dislocations in pristine Fe without H was determined by:

$$E(R) = \frac{1}{L} \sum_{r_i < R} \left( \overline{E_{i,Fe}^{dis}} - \overline{E_{Fe}^{0}} \right), \tag{2}$$

where $L$ is the dislocation line length along the longitudinal axis, $\overline{E_{i,Fe}^{dis}}$ is the average potential energy of the $i^{th}$ Fe atom within the region bounded by the cutoff radius $R$, and $\overline{E_{Fe}^{0}}$ is the average potential energy of a Fe atom in a reference perfect bcc lattice at 300 K.

The influence of H on dislocation was quantified by calculating the binding free energy of H to different dislocations, i.e., the change in free energy due to H binding:

$$F_b^H = \bar{n}\mu^H - (\overline{E_{tot}^{dis+H_n}} - \overline{E_{tot}^{dis}}), \tag{3}$$

where $\bar{n}$ is the average number of H, $\mu^H$ is H chemical potential that corresponds to the given H concentration, $\overline{E_{tot}^{dis+H_n}}$ and $\overline{E_{tot}^{dis}}$ respectively denotes the average of total potential energy of the system with and without H atoms. Note here we use $\overline{E_{tot}^{dis+H_n}} - \overline{E_{tot}^{dis}}$ to represent H free energy in the dislocation system, namely, neglecting entropy effects since H clusters show an ordered rock-salt structure around dislocations (see Refs. [22,23]).

**Simulations of dislocation-dislocation reaction.** An orthogonal simulation cell was employed with basis vectors along $x \parallel [001]$, $y \parallel [1\bar{1}0]$, $z \parallel [110]$, and a size of $202 \times 242 \times 143$ Å$^3$ (see supplementary for further details). Atomistic models containing multiple dislocations were constructed following the method introduced by Zhang et al. [52]. In simulations without pre-existing dislocation loops, two <111>/2 screw dislocations were inserted on neighboring $(1\bar{1}0)$ planes located at 1/4 of the y axis. For simulations involving pre-existing dislocation loops, a rectangular vacancy-type loop with a side length of 40 Å was introduced, positioned on the y+ side of the intersection of the two <111>/2 screw dislocations. The loop was placed on the (001) plane, with a Burgers vector of [001] or [002]. Free surface boundary conditions were applied in all three directions, and atoms in the outer shell of the simulation box, with a thickness of 7 Å, were held fixed. In simulations involving shear loading, a shear strain was applied by displacing the atoms in the fixed shell along the designated shear direction, with a strain rate of 0.05/ns.

**Data availability**

The data generated and/or analyzed within the current study will be made available upon reasonable request to the authors.

**Competing interests**

The authors declare no competing interests.

# Supplementary information for modulating dislocation reactions through preferential hydrogen segregation in bcc metals

**S1. Accuracy of the interatomic potential in predicting core structure and H binding for different dislocations**

Here we present additional simulations to validate the accuracy of the Fe-H interatomic potential [1,2] for predicting dislocation core structures and single H binding energies. Fig. S1 As demonstrated in considered here show compact core with relatively small spreading in the slip plane and do not dissociate into partial dislocations, agreeing well with those reported in previous DFT studies [3-5]. Using these structures, we mapped the binding energy of single H atoms around dislocations by inserting H into tetrahedral and octahedral interstitial sites. The results, illustrated in Figs. S1 and S2, reveal that edge dislocations exhibit stronger H binding than screw dislocations, with maximum binding energies closely matching DFT predictions (0.47 eV for <111>/2 edge dislocations, 0.256–0.27 eV for <111>/2 screw dislocations [3,4], no DFT data available for H interaction with <001> dislocations to our best knowledge). Notably, we observed significant long-range elastic interactions between H and <001> edge, <111>/2 edge, and <111>/2 screw dislocations, while such interactions were absent for the <001> screw dislocation, consistent with predictions from linear elasticity theory [3-6].

Notably, in our previous study [7], we demonstrated that H clustering under varying stress states can be accurately captured by the interatomic potential used here, ensuring a reliable description of long-range elastic interactions between H and dislocations. Given the strong agreement with both dislocation core structures and H binding energies reported in DFT studies, we are confident that this interatomic potential is well-suited for investigating H effects on dislocation reactions.

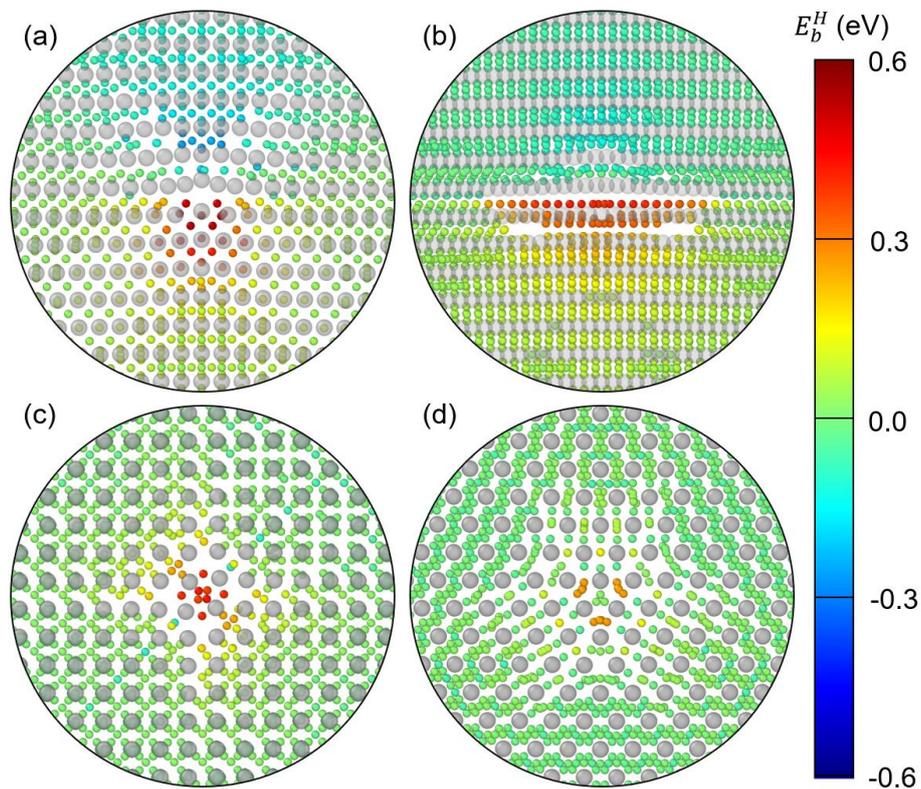

Figure S1. Binding energy map of a single H atom around (a) edge {110}⟨001⟩, (b) edge {1$\bar{1}$0}⟨111⟩/2, (c) screw ⟨001⟩, and (d) screw ⟨111⟩/2 dislocations. Large grey spheres indicate the positions of Fe atoms in a pristine dislocation model, while small spheres represent the final positions of H atoms.

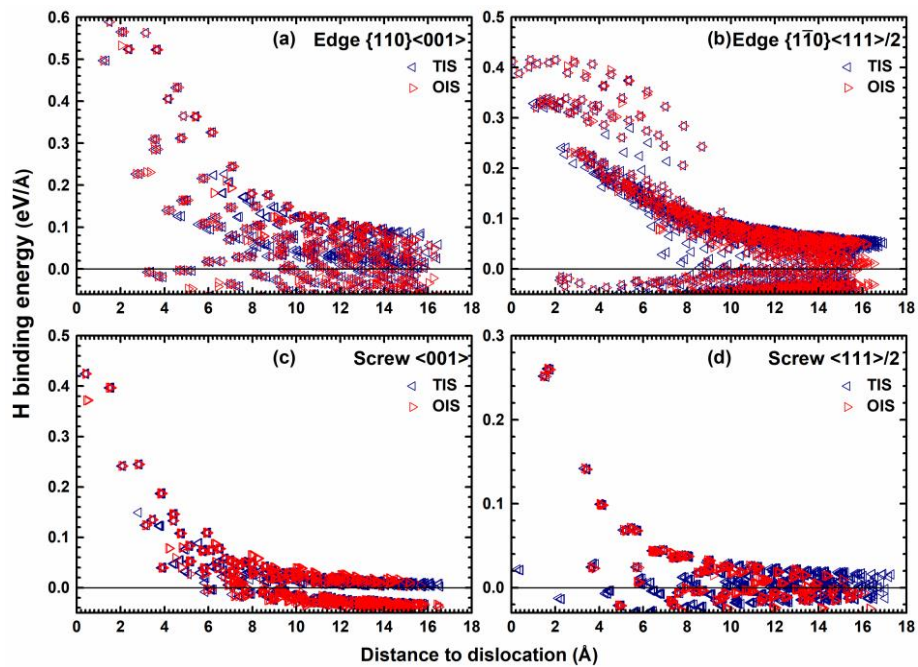

Figure S2. Binding energy of a single H atom around (a) {110}⟨001⟩ edge, (b) {1$\bar{1}$0}⟨111⟩/2 edge, (c) ⟨001⟩ screw, and (d) ⟨111⟩/2 screw dislocations. Left and right triangles indicate data calculated with H initially placed at tetrahedral (TIS) and octahedral (OIS) interstitial sites, respectively.

**S2. Type of dislocation junctions formed from two <111>/2 screw dislocations**

Since screw dislocations do not have a defined slip plane, two intersecting screw dislocations can zip together along various directions to form a binary junction. However, it is reasonable to assume that they will zip along their intersecting plane and bisecting direction in order to minimize dislocation length and, consequently, line energy. In bcc metals, two <111>/2 screw dislocations always intersect along a {110} plane while bisect along the <001> and <110> directions (see Fig. S3). It is important to note that the sign of a dislocation's Burgers vector is direction-dependent. If the Burgers circuit is drawn from the opposite direction, the Burgers vector will reverse.

Thus, when two ⟨111⟩/2 screw dislocations zip along different directions, they may form junctions with different Burgers vectors. The above is further illustrated in Fig. S3, where two different orientation of screw dislocations are considered: one with $\vec{b_1} = [111]/2$ and $\vec{b_2} = [\bar{1}\bar{1}1]/2$ (**orientation I**), and the other with $\vec{b_1}$ unchanged but $\vec{b_2} = [11\bar{1}]/2$ (**orientation II**). Any other orientation of two intersecting screw dislocations will be symmetrically equivalent to one of these two cases.

For orientation I (see Fig. S3a), zipping along [110] direction produces a [001] edge junction, while zipping along [001] direction produces a [110] edge junction. For orientation II (see Fig. S3b), zipping along [110] direction results in a [110] screw junction, while zipping along [001] direction in turn forms a [001] screw junction. Considering that <110> dislocations are generally unstable in bcc metals and thus can be excluded as possible dislocation products, we expect that orientation I and II will lead to the formation of <001> edge and screw dislocation junctions, respectively. Since the H binding effect with ⟨001⟩ screw dislocations is relatively weak (cf. Fig. 2 in the main text), we anticipate that H will not significantly impact the dislocation reaction in Orientation II (see additional discussion in Section S4). Therefore, in this work, we primarily focus on the formation of ⟨001⟩ edge dislocation junctions.

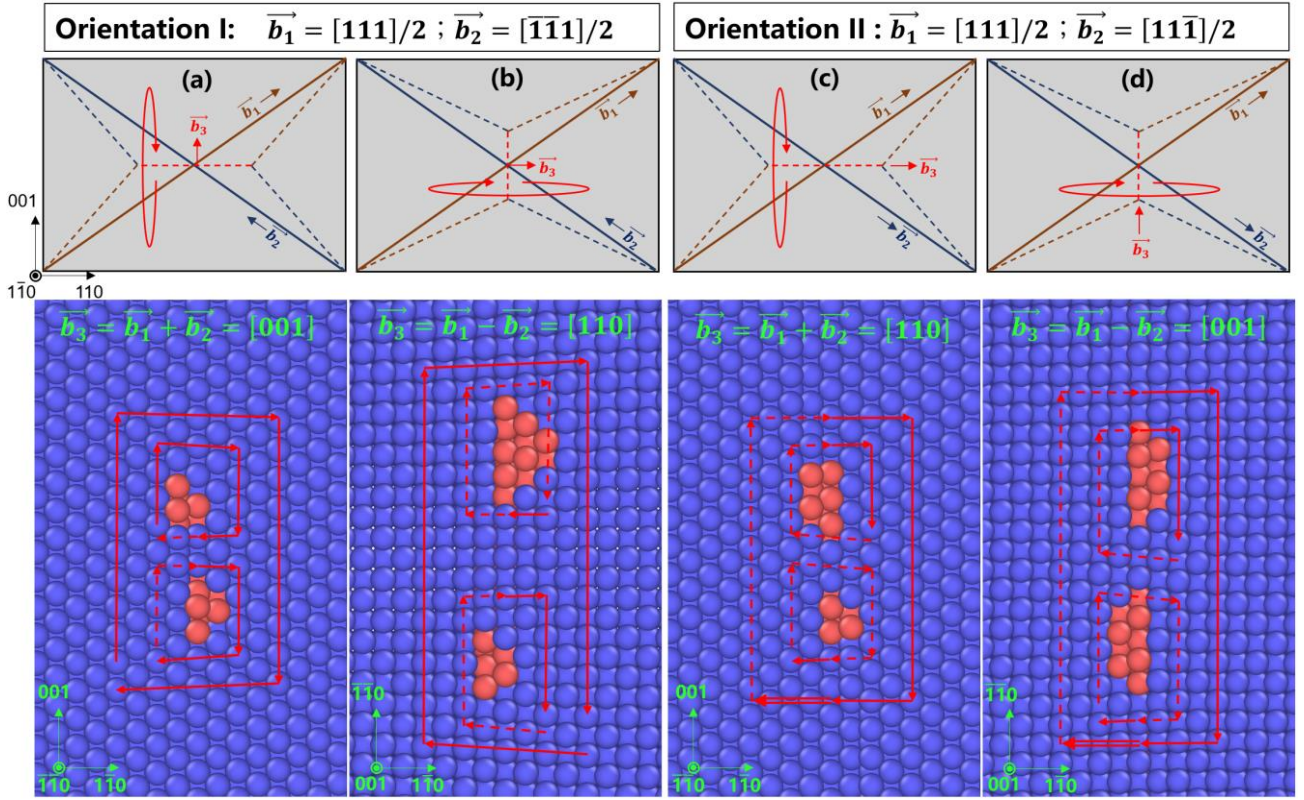

Figure S3. Burgers vector analysis for two intersecting <111>/2 screw dislocations in bcc lattice. (a) and (b) show junctions zipped along [110] and [001] directions, respectively, for Orientation I. (c) and (d) illustrates the same for orientation II. In the top panels, solid and dotted lines represent dislocations before and after junction formation, respectively, with red circles indicating the Burgers vector analysis circuit. The bottom panels display corresponding atomistic structures on the Burgers circuit plane. Blue and orange spheres represent Fe atoms with bcc and non-bcc structures, respectively. Solid lines represent Burgers segments on the visible atomic layer, dotted lines denote segments below the visible layer, while double lines indicate segments above the visible layer.

**S3. Line tension model for junction formation in general bcc metals**

The line energy of a dislocation of type *i* can be expressed as [8]:

$$E_i(R) = K_i \ln\left(\frac{R}{r^c}\right) + E_i^c, \qquad (S1)$$

where $R$ is the cut-off radius for evaluating the line energy, $K_i$ is a pre-logarithm constant, $r^c$ represent the radius of a highly distorted dislocation core region unsuitable for linear elastic theory description, and $E_i^c$ denotes the energy of the dislocation core (typically a small value). Using the line energy data, the energy change $\Delta E$ for the formation of a <001> edge/screw dislocation junction can be described by the line tension model (cf. Eq. 1 in the main text):

$$\Delta E = E_3 - (E_1 \cos\theta_1 + E_2 \cos\theta_2). \qquad (S2)$$

where $E_1 = E_2$ represents the line energy for the <111>/2 screw dislocation, and $E_3$ is the line energy for the <001> edge or screw dislocation. For the formation of a <001> edge dislocation junction, $\theta_1 = \theta_2 = 35.3°$. For the formation of a <001> screw dislocation junction, $\theta_1 = \theta_2 = 54.7°$.

Using anisotropic linear elastic theory [9] and elastic constants reported from previous studies [10-14], we determined the pre-logarithm constant $K_i$ for screw and edge type <111>/2 and <001> dislocations in Fe, W, Mo, Cr, Ta, Nb, and V. By substituting $R = 200$ Å, and assuming $E_i^c = 0$ at $r^c = 1$ Å in Eq. S1, we obtained line energies for these dislocations. As listed in Table S1, the line energies predicted by anisotropic linear elastic theory align well with our MD predictions for Fe and W, confirming the accuracy of the results. Using these line energy data, we calculated the $\Delta E$ for the formation of <001> edge and screw junctions, with results also presented in Table S1. Here we find that all $\Delta E_{edge}$ values are positive, indicating that the formation of a <001> edge dislocation junction from two <111>/2 screw dislocations is energetically unfavorable across all bcc metals considered. For the case of <001> screw dislocation junction, we observed positive $\Delta E_{screw}$ values in Fe, W, Ta, while negative values were found in Mo, Cr, Nb, V, consistent with the findings reported in Ref. [15].

Table S1. Line energies of <111>/2 and <001> dislocations in bcc metals, calculated using anisotropic linear elastic theory and MD simulations (for Fe and W). Corresponding energy changes for <001> edge and screw junction formation are also shown. All values are in eV/ Å.

|  |  | Edge $\{1\bar{1}0\}\langle001\rangle$ | Edge $\{1\bar{1}0\}\langle111\rangle/2$ | Screw $\langle001\rangle$ | Screw $\langle111\rangle/2$ | $\Delta E_{edge}$ | $\Delta E_{screw}$ |
|---|---|---|---|---|---|---|---|
| MD | Fe | 2.44 | 2.01 | 2.49 | 1.17 | 0.53 | 1.14 |
|  | W | 5.81 | 4.42 | 4.18 | 3.11 | 0.72 | 0.59 |
| Elastic theory | Fe | 2.44 | 2.03 | 2.51 | 1.16 | 0.54 | 1.18 |
|  | W | 5.76 | 4.32 | 4.14 | 3.10 | 0.70 | 0.55 |
|  | Mo | 4.65 | 3.41 | 2.89 | 2.61 | 0.38 | -0.12 |
|  | Cr | 3.47 | 2.51 | 2.20 | 2.11 | 0.03 | -0.23 |
|  | Ta | 3.02 | 2.36 | 2.42 | 1.38 | 0.77 | 0.83 |
|  | Nb | 1.82 | 1.33 | 0.80 | 1.02 | 0.16 | -0.37 |
|  | V | 1.80 | 1.33 | 1.03 | 0.92 | 0.30 | -0.03 |

## S4. Effect of H on the formation of <001> screw dislocation junction

This section presents additional results for the formation of <001> screw dislocation junction from two <111>/2 screw dislocations under varying H concentrations. We employed modeling procedures analogous to those described in the main text but began with two <111>/2 screw dislocations oriented according to orientation II. According to the discussions in Section S2, this orientation is expected to lead to the formation of <001> screw dislocations, if they were to form at all. The simulation results are illustrated in Fig. S4. Our findings indicate that the two dislocations persist in a bounded crossed state across all explored H concentrations, including the H-free case (not shown). Importantly, no evidence of <001> dislocation junction formation was observed within the range of H concentrations examined. This outcome corroborates our initial hypothesis that H does not facilitate the formation of <001> screw dislocation junctions.

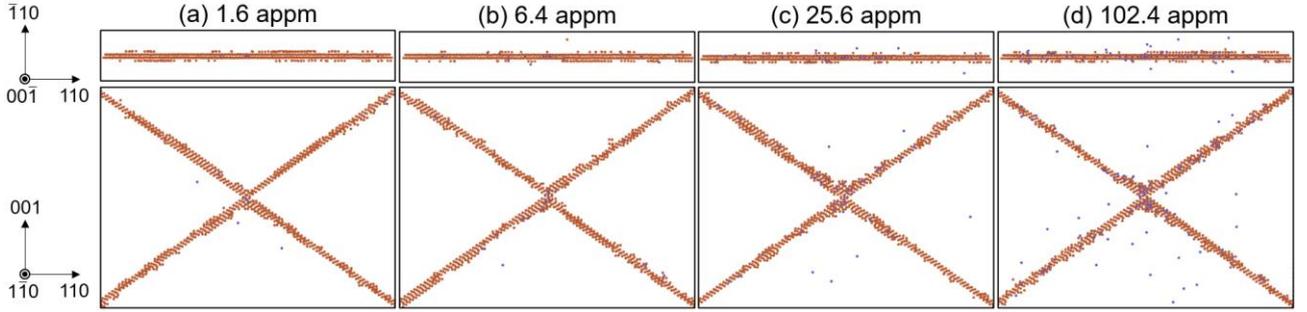

Figure S4. Reaction of two <111>/2 screw dislocations in orientation II. Orange atoms are Fe (only non-bcc Fe atoms are visualized here) and blue ones are H. (a-d) show results with different far-field H concentrations. Top panels show top views from $[00\bar{1}]$ direction under different far-field H concentrations. Bottom panels are corresponding front views from $[1\bar{1}0]$ direction.

**S5. Additional details for simulation setups**

This section provides further information and visualizations regarding the atomistic models and simulation setups used in this work. Fig. S5a illustrates the cylindrical model employed in simulations of individual straight dislocations. Fig. S5b depicts the model used to simulate the interaction between two <111>/2 screw dislocations. The movements of these two dislocations are controlled by applying shear in the XZ plane at an angle θ\theta θ to the X-axis. For a given shear strain value, changing $\theta$ results in a variation in the effective shear acting on the two dislocations. This can be quantified by calculating the resolved shear strain using 2-dimensional tensor rotation:

$$\begin{pmatrix} 0 & 0 & \varepsilon \\ 0 & 0 & 0 \\ \varepsilon & 0 & 0 \end{pmatrix} \xrightarrow{\text{rotate } \varphi} \begin{pmatrix} \varepsilon \sin 2\varphi & 0 & \varepsilon \cos 2\varphi \\ 0 & 0 & 0 \\ \varepsilon \cos 2\varphi & 0 & -\varepsilon \sin 2\varphi \end{pmatrix}. \qquad (S1)$$

where $\varphi$ represent the angle between the dislocation Burgers vector and the shear direction, and $\cos 2\varphi$ represent the ratio of resolved shear strain along the dislocation. For two <111>/2 dislocations with orientation I, we have $\varphi_1 = \theta - 35.26°$ and $\varphi_2 = \theta + 125.26°$. While for two <111>/2 dislocations with orientation II, the angles are $\varphi_1 = \theta - 35.26°$ and $\varphi_2 = \theta + 35.26°$.

Fig. S3 shows the resolved shear ratio for two <111>/2 dislocations of orientation I at varying $\theta$. Two specific values of $\theta$ were employed in this work:

a) In simulations of H-enabled <001> edge dislocation junction formation, a shear load with $\theta = 0°$ was applied, which exerts opposing forces on the two dislocations, driving them move towards each other. The simulation procedure involved the following steps: i) Slowly increasing the shear strain to approximately 0.45%, under which the two dislocations form a metastable structure known as the crossed state; ii) Maintaining the strain level while activating GCMC H charging until the system reaches equilibrium; iii) Unloading the system to a zero strain state while continuing H charging until a new equilibrium is established.

b) In simulations of H induced <001> vacancy loop formation and accumulation, a shear load with $\theta = 45°$ was adopted which exerts equal forces on both <111>/2 dislocations, driving them to move in the $[1\bar{1}0]$ direction.

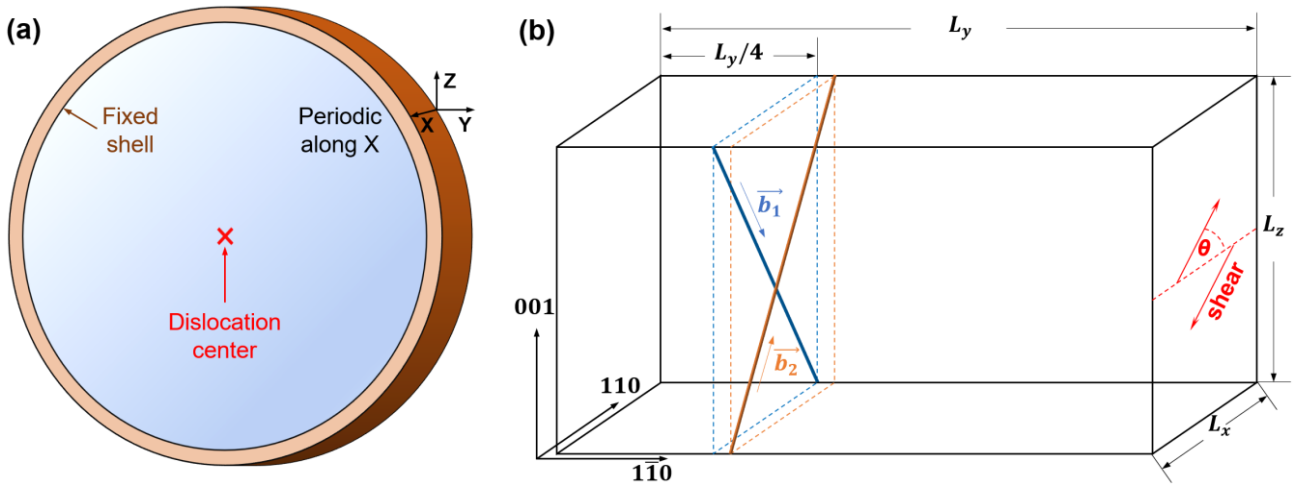

Figure S5. Schematic representations of the atomistic models used in simulations. (a) The model for simulating a straight dislocation line, featuring a cylindrical geometry with a thickness of approximately 20 Å along the X-axis and a radius of 150 Å. The outer shell, which has a radial thickness of 7 Å, remains fixed throughout all simulations. (b) The model for simulating the interaction between two <111>/2 screw dislocations, with dimensions $L_x = 202$ Å, $L_y = 242$ Å, and $L_z = 143$ Å. In this setup, the two screw dislocations are separated by one atomic layer along the $[1\bar{1}0]$ direction. The outer shell of the box, also with a thickness of 7 Å (not shown), is held fixed and only responds to the applied shear strain.

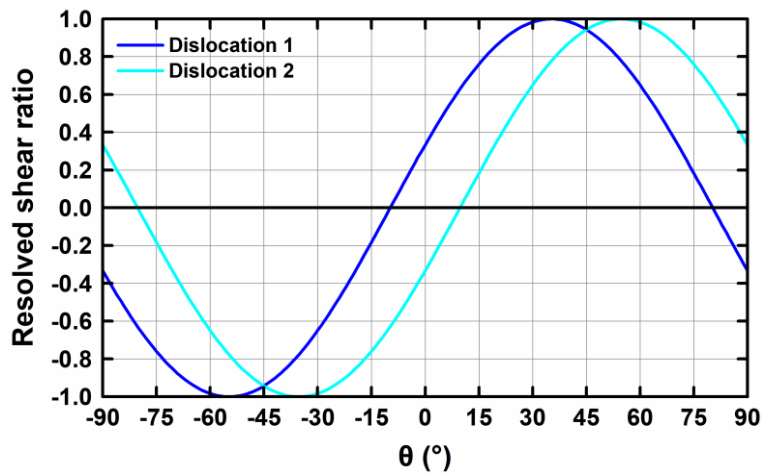

Figure S6. Ratio of resolved shear strain for two <111>/2 screw dislocations with orientation I. The angle $\theta$ denotes the angle between shear direction and the X axis, as referenced in Fig. S5.

## S6. Stress fields around <001> edge junction segments of different lengths

Fig. S7 shows [001] tensile stress profile at the center of [001] edge dislocation junction segments of varying lengths. In this simulation, no H was introduced; thus, atoms near the junction nodes were fixed to prevent spontaneous unzipping. As illustrated, shorter junction segments exhibit smaller tensile stress fields, leading to a reduced capacity to trap H and H clusters. This observation accounts for the relatively small sizes of H clusters shown in Figs. 3c-4d in the main text.

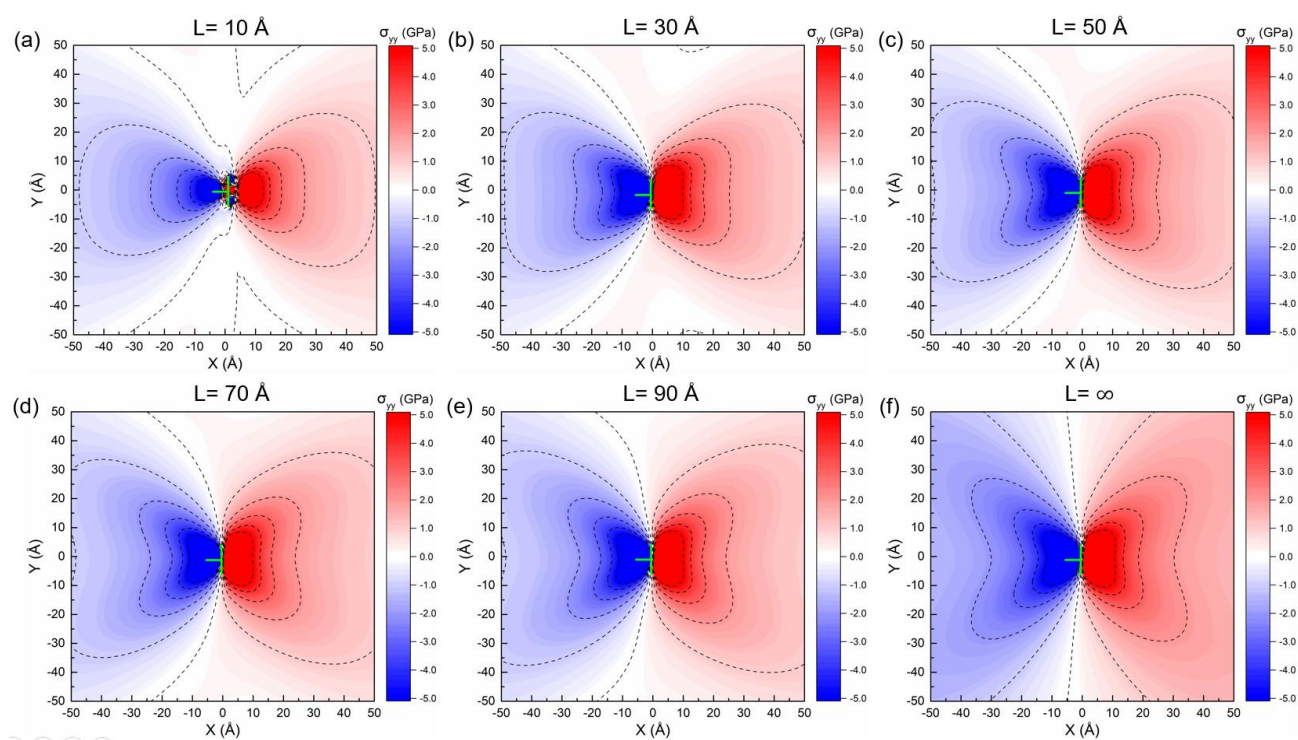

Figure S7. [001] tensile stress profile at the center of [001] edge dislocation junction segments of different lengths. $L = \infty$ represents an infinite long dislocation line.

## S7. Preliminary results for vacancy loop open up

We demonstrated that the dislocation reaction under H charging can facilitate the formation and accumulation of vacancy-type dislocation loops. It is reasonable to anticipate that these vacancy loops may further expand within the bulk lattice, potentially evolving into H bubbles or blisters. However, the EAM type interatomic potential employed in this study was specifically formulated to describe H behavior in solid iron and does not accurately represent H interactions in open spaces such as cracks or blisters [1,2]. To explore the possible evolution of these vacancy loops, we conducted additional MD simulations without the presence of H. Beginning with the <002> vacancy loop structure illustrated in Figs. 4i-4j, we systematically removed all H atoms from the system and performed MD simulations without H charging. As depicted in Fig. S8, the <002> vacancy loop partially opened after 1 ns of MD simulation, transforming into a combination of a <001> vacancy loop and a small crack oriented along the (001) plane.

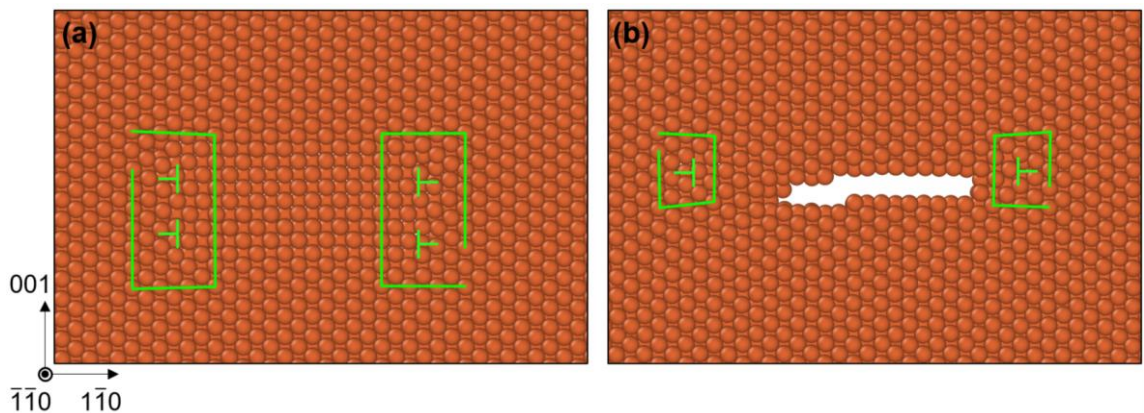

Figure S8. Evolution of the <002> Vacancy Loop. (a) Initial configuration at $t = 0\ ns$ and (b) subsequent configuration at $t = 1\ ns$ following H removal. Only the atoms near the center cross-section of the vacancy loop are visualized for clarity.

**S8. Preliminary results for W-H system**

Given the preferential hydrogen clustering observed under <001> tensile stress across various bcc metals (including Fe, W, Mo, and Cr) [7], we conducted preliminary simulations utilizing a W-H interatomic potential [16] that suitable for modelling H-H clustering (similar potentials are not available for Mo-H and Cr-H system to our best knowledge). The results, presented in Fig. S9, indicate that two <111>/2 screw dislocations initially bow out from each other at low H concentrations (or without H), demonstrating a repulsive interaction. However, as the hydrogen concentration increases to $C_H \geq 10^{-12}$, a <001> edge junction segment begins to form and expand from the two <111>/2 screw dislocations, accompanied by a significant accumulation of H at the tension side of the junction. In general, these results are quite similar to that in Fe, except the H concentration required for notable junction formation is considerably lower in W ($10^{-12}$ in W against $2.56 \times 10^{-5}$ in Fe). This variation can be attributed to differences in H-H binding in two systems, with the energy released during H clustering being approximately 0.69 eV/H in W, nearly double the value in Fe (0.32 eV/H)[7]. Consequently, preferential H clustering and the formation of H-enabled <001> edge junctions can occur at significantly lower H concentrations in W.

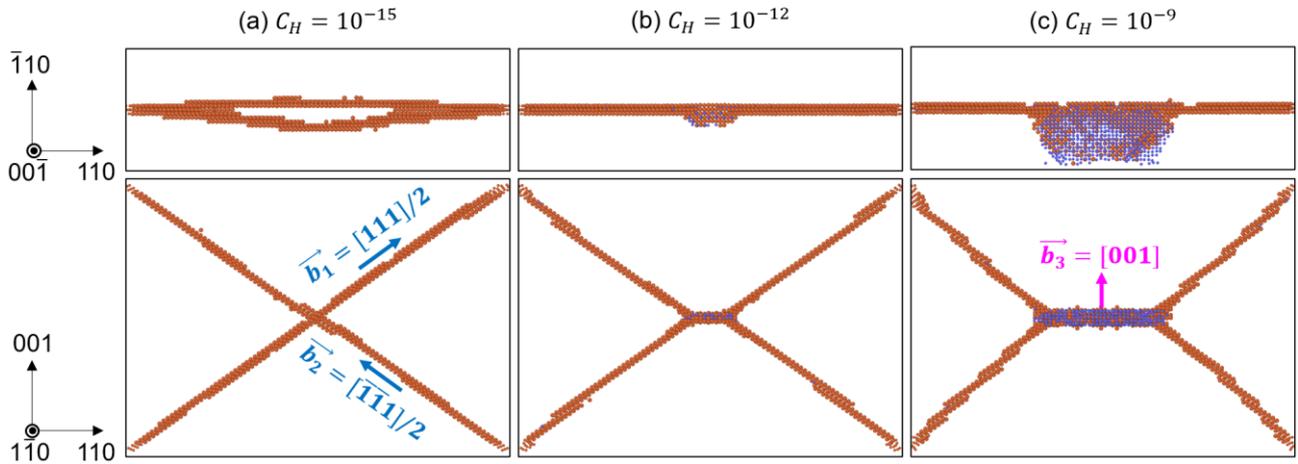

Figure S9. Reaction of two <111>/2 screw dislocations with orientation I in bcc W. Orange atoms are W (only non-bcc W atoms are visualized here) and blue ones are H. (a-c) show results with different far-field H concentrations. Top panels show top views from $[00\bar{1}]$ direction under different far-field H concentrations. Bottom panels are corresponding front views from $[1\bar{1}0]$ direction.

**S9. Stress-strain curve with/without H**

Fig. S9 presents the stress-strain curve during the sliding of two <111>/2 screw dislocations with orientation I under a concentration of 102.4 appm H (i.e., the simulation shown in Fig. 4a-5e). For comparison, the results from the H-free scenario are also plotted. Notably, the yield stress in the H-free case, while not precisely defined due to the continuous bowing of dislocations, approximates half that observed in the H-charged condition. This increase in yield stress induced by H is attributed to the formation of dislocation junctions, which serve as effective pinning sites for the two <111>/2 screw dislocations. Consequently, each <111>/2 screw dislocation segment is effectively separated into two individual segments under H charging, thereby increasing the difficulty of dislocation bowing and effectively doubling the yield stress.

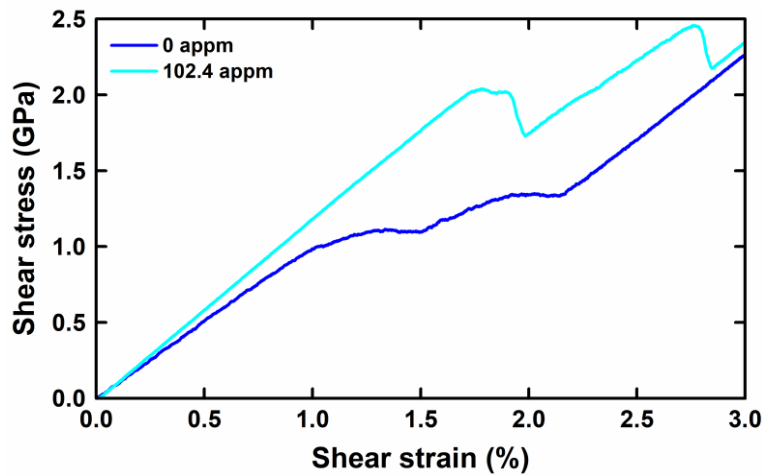

Figure S10. Stress-strain curve during the sliding of two <111>/2 screw dislocations with orientation I under 0 or 102.4 appm of H.